# A nudge to the truth: atom conservation as a hard constraint in models of atmospheric composition using an uncertainty-weighted correction

Patrick Obin Sturm[a] and Sam J. Silva[a,b,c]

AUTHOR ADDRESS

[a]Department of Earth Sciences, University of Southern California, Los Angeles, CA

[b]Department of Environmental Engineering, University of Southern California, Los Angeles, CA

[c]Department of Population and Public Health Sciences, University of Southern California, Los Angeles, CA

**ABSTRACT:** Computational models of atmospheric composition are not always physically consistent. For example, not all models respect fundamental conservation laws such as conservation of atoms in an interconnected chemical system. In well performing models, these nonphysical deviations are often ignored because they are frequently minor, and thus only need a small nudge to perfectly conserve mass. Here we introduce a method that anchors a prediction from any numerical model to physically consistent hard constraints, nudging concentrations to the nearest solution that respects the conservation laws. This closed-form model-agnostic correction uses a single matrix operation to minimally perturb the predicted concentrations to ensure that atoms are conserved to machine precision. To demonstrate this approach, we train a gradient boosting decision tree ensemble to emulate a small reference model of ozone photochemistry and test the effect of the correction on accurate but non-conservative predictions. The nudging approach minimally perturbs the already well-predicted results for most species, but decreases the accuracy of important oxidants, including radicals. We develop a weighted extension of this nudging approach that considers the uncertainty and magnitude of each species in the correction. This species-level weighting approach is essential to accurately predict important low concentration species such as radicals. We find that applying the uncertainty-weighted correction to the nonphysical predictions slightly improves overall accuracy, by nudging the predictions to a more likely mass-conserving solution.

**SYNOPSIS:** Computational models of atmospheric composition don't always make scientifically trustworthy predictions. This corrective approach nudges the predicted concentrations of chemical species based on their uncertainties to guarantee that predictions respect conservation of mass.

## 1 INTRODUCTION

Computational models of atmospheric chemistry are fundamental tools used to support basic research, inform environmental policies and assessments, and predict future atmospheric composition. However, these numerical models can sometimes violate core tenets of our understanding of the natural world, which includes the principle of conservation of mass, or atoms in the case of chemical reactions. This is often caused by parameterization choices, for example from incomplete reaction networks in chemical mechanisms[1–3] or transport in 3D air quality models[4,5]. These violations are particularly common when using data-driven methods for parameterization which are not based solely on first principles like many process-based models, but rather on a set of parameters optimized on some given data.

To address this problem, approaches for ensuring physical constraints in data-driven models have been developed for several applications in the Earth system sciences[6–10]. For example, Geiss et al.[7] constrain convolutional neural networks predicting the mixing ratios of atmospheric chemical species, to ensure the average of spatially super-resolved mixing ratios is consistent with the mixing ratio at their corresponding coarse grid cell, done analogously for conservation of energy or atmospheric water mass in Harder et al.[8] Harder et al.[10] address mass violation in aerosol fields by adjusting the variable with the least accurate predictions to satisfy mass conservation; this is related to a general approach proposed by Beucler et al.[6] which constrains a subset of the output by the input and another subset of the output that is unconstrained. In another aerosol application, Sturm et al.[9] use scaling factors to ensure that a set of advected superspecies conserves total concentration in both particle and gas phases. However, none of these methods can be used to conserve atoms in data-driven models of atmospheric chemistry, which has additional complexity arising from molecular composition. For example, a scaling factor for each element cannot be used to conserve atoms, as molecular species are formed by mixtures of atoms from these different elements.

The mass-conserving framework for atmospheric chemistry introduced in Sturm and Wexler[11], based on relating atom fluxes to chemical tendencies, can be used to enforce atom conservation as a hard constraint for machine learning models. This framework serves as a machine-learning analog to traditional finite volume numerical methods[11–14]. However, for atom conservation this framework requires either specialized data on chemical fluxes or custom graph neural network architectures[15–19] embedded with a bipartite network relating reactions to species[20]. Recent work by Döppel and Votsmeier[21] point out that the detailed stoichiometry and reaction network connectivity is not known in some contexts. Requirement of specialized training data on fluxes, machine learning architectures, or knowledge of the reaction network could limit the more widespread adoption of this mass-conserving framework. For example, this framework is not directly compatible with physics-based models or other machine learning methods for predicting atmospheric composition like random forests[22] or XGBoost[23] emulating the GEOS-Chem mechanism[24], where every species has its own dedicated machine learning model trained to predict concentration, bias, or tendency. Such models require a different strategy for enforcing conservation.

This work introduces a corrective approach that minimally adjusts the predicted concentrations of chemical species to satisfy a set of elemental conservation laws exactly. This model-agnostic approach uses a single matrix multiplication step to nudge predictions to the closest physically consistent solution. We demonstrate this corrective approach on the gradient-boosting[25] machine learning algorithm XGBoost[26] that is trained to emulate a reference model of ozone formation[11]. We derive a weighted version of this nudging approach informed by species-level uncertainty, preferentially adjusting species with lower accuracy and larger changes to meet the strict elemental conservation laws.

## 2 METHODS AND MODEL DESCRIPTION

### 2.1 Nudging tendencies for conservation during incremental change

The following section lays out a mathematical framework for ensuring any numerical prediction conserves atoms. In a chemical system with $m$ compounds formed by $p$ elements, a vector of species tendencies $\Delta C' \in \mathbb{R}^m$ represents the concentration change of each species over a given timestep. The atomic composition of each species is defined by composition $M \in \mathbb{Z}_{\geq 0}^{m,p}$. If atom conservation is not explicitly specified within a particular predictive model, then conservation of atoms might be violated in predictions, stated mathematically:

$$M^T \Delta C' \neq 0_p \tag{1}$$

Where $0_p$ is a $p$-element zero vector. This means that that the total number of atoms for one or more elements is changed by this prediction: the prediction artificially removed mass from or introduced mass to the system in a nonphysical way.

We seek a way to minimally nudge $\Delta C'$ to $\Delta C$ such that atoms are conserved. Mathematically, this means the constraint $M^T \Delta C = 0_p$ is satisfied. We can recognize this as a constrained optimization problem, where we want to minimize the difference between nudged solution $\Delta C$ and the original prediction $\Delta C'$. If we define minimal to be with respect to the squared difference between these two predictions, this optimization represented mathematically is $min_{\Delta C} \|\Delta C' - \Delta C\|_2$. This becomes a constrained least-squares problem with a closed-form solution[27]. The minimally nudged mass-conserving $\Delta C$ to any nonphysical $\Delta C'$ can be obtained by a single matrix operation:

$$\Delta C = M_{fix} \Delta C' \tag{2}$$

Where the correction matrix $M_{fix} \in \mathbb{R}^{m,m}$ that projects any $\Delta C'$ onto the closest mass-conserving $\Delta C$ is defined by

$$M_{fix} = (I - M(M^T M)^{-1} M^T) \tag{3}$$

The derivation for this closed-form solution is shown in the supplement S1 as a special case of the more general weighted least squares optimization problem (section 2.2). Note that the constraints on positivity are removed for $M_{fix}$, due to the subtraction from $I$ and the use of the Moore-Penrose pseudoinverse $M^+ = (M^T M)^{-1} M^T$. The Moore-Penrose pseudoinverse itself can give positive or negative values: negative values are nonphysical for some applications such as extents of forward reaction[11] or superspecies concentrations[9]. Negative tendencies that satisfy $-M_{fix} \Delta C' < C$ elementwise are still valid (though tendencies without flux constraints can sometimes lead to negative concentrations[19]. This method could also be modified to nudge absolute concentrations rather than incremental changes (tendencies): this approach is detailed in the supporting information section S2.

We note that the composition matrix $M$ is used to ensure atom conservation in another related machine learning application, where the null space of $M^T$ is used as a balance layer in an atom-conserving chemical reaction neural network used to discover stoichiometries in reaction systems[21]. Analogously, the corrective approach in Equation 3 nudges $\Delta C'$ to be in the null space of $M^T$, acting on concentrations instead of stoichiometric weights. The approach as tested here is a corrective nudge on predictions, rather than a built-in deep learning constraint, though it could also be built into neural networks for atom conservation in future applications.

Though this constrained least-squares nudging approach is straightforward and computationally feasible in its closed form, it is sparsely used in related literature in the atmospheric and chemical sciences. Least squares methods are sometimes used in solving the Stokes equations for fluid flow problems, but some of these approaches only conserve mass in the "weak sense" as part of the minimization problem rather than the hard constraint[28,29]. Such soft constraint approaches are analogous to the more recent physics-informed neural networks[30], which encourage adherence to conservation laws or other known physics[6,31] as regularization terms in the training loss function, often minimized with respect to squared error. Projection-based approaches for hard constraints in neural networks have been developed for more general applications outside of the atmospheric sciences[32,33]. In the atmospheric sciences, Brown[34] developed a least-squares nudging approach for mass conservation as a hard constraint in collisional breakup of raindrops, for which mechanisms and appropriately mass-conserving numerical methods are still being developed[35]. Brown[34] found that this least-squares nudging approach improved model comparisons to droplet spectra compared to using scaling factors for mass conservation. While we develop this nudging approach for atom conservation in at-mospheric chemistry models, we note that this is readily generalizable to other applications in the intersection of physical sciences and data-driven numerical methods.

### 2.2 Species-level weights to factor in scale and uncertainty

In the atmospheric chemical system, species span a wide range of concentrations, lifetimes, uncertainty, and societal importance. We adjust the method in section 2.1 to assign *species-level* weights in the optimization equation (with the hard constraint $M^T \Delta C = 0_p$ unchanged),

$$min_{\Delta C} \|W(\Delta C' - \Delta C)\|_2 \quad (4)$$

where $W \in \mathbb{R}_{\geq 0}^{m,m}$ scales the adjustment of each species by some measure of its importance. The higher the species-level weight, the less it is adjusted in the correction to atom conservation. With this adjustment, the weighted correction matrix becomes

$$M_{fix,weighted} = I - (W^T W)^{-1} M (M^T (W^T W)^{-1} M)^{-1} \quad (5)$$

In the case that $W$ is identity, or a diagonal matrix with equal weights, the problem simplifies to the unweighted case in section 2.1. The derivation of the weighted correction approach using Lagrange multipliers, and the special case of the unweighted approach, is shown in section S1 in the supporting information.

The choice of the species-level weighting is flexible as long as the weights remain positive, and could be hand selected using domain knowledge, scaled by the relative tendencies of the species, or determined from a measure of uncertainty the species prediction. As a proof of concept, we consider both the magnitude of tendencies and the uncertainty in prediction with the species-level weight:

$$w_i = \frac{1}{\left(1 - ReLU(R_i^2)\right)\overline{|\Delta C_i|}} \quad (6)$$

Where $\overline{|\Delta C_i|}$ is just a scaling factor equal to the average absolute tendency of species $i$, to account for the wide spread in magnitude of tendencies. Species-level uncertainty is represented by the $1 - ReLU(R_i^2)$ factor, $R_i^2$ being the coefficient of determination between the predicted tendency and the truth. $ReLU(x)$ is the rectified linear unit function, defined as $ReLU(x) \stackrel{\text{def}}{=} \max(0, x)$ to ensure the positivity constraint on $w_i$ if species $i$ is very poorly predicted. The species-level weights $w_i$ compose the values of the diagonal matrix $W$.

On a scaled basis using $\overline{|\Delta C_i|}$, we note that weighted approach converges to an atom-conserving extension of the completion method in Harder et al.[10], in the event that one variable is by far the worst performing or if the species-level weight is made to be more sensitive to uncertainty than in Equation 6.

### 2.3 Example using the primary photolytic cycle

We use a fundamental cycle in atmospheric chemistry to demonstrate the projection approach. The primary photolytic cycle between nitrogen oxides and ozone in the troposphere, from which the Leighton relationship can be derived, is shown in Table 1.

| Table 1: Primary photolytic cycle | |
|---|---|
| Reaction | Reaction Number |
| $NO_2 + hv \rightarrow NO + O$ | R1 |
| $O + O_2 \rightarrow O_3$ | R2 |
| $O_3 + NO \rightarrow NO_2 + O_2$ | R3 |

The matrix $M \in \mathbb{R}_{\geq 0}^{m,p}$ is a composition matrix of $m$ compounds formed by $p$ elements, $m = 5$ and $p = 2$ in this example, and

$$M = \begin{bmatrix} species & \text{N} & \text{O} \\ \text{O}_3 & 0 & 3 \\ \text{NO} & 1 & 1 \\ \text{NO}_2 & 1 & 2 \\ \text{O} & 0 & 1 \\ \text{O}_2 & 0 & 2 \end{bmatrix}$$

We demonstrate the following example of both nudging approaches, visualized in Figure 1. In this example, $\Delta C \in \mathbb{R}_{\geq 0}^{5,1}$ is a vector of concentrations or mixing ratios for the 5 different molecular species: if R1 proceeds at 5 ppb/min, R2 at 4 ppb/min, and R3 at 2 ppb/min, then the correct $\Delta C = [2, 3, -3, 1, -2]$ in ppb, as shown by the red circle in Figure 1. If instead an incorrect nonphysical vector $\Delta C' = [2.00, 3.00, -2.00, 1.02, -2.20]$ is predicted as shown by the blue triangle, the prediction is moved off of the physically possible constrained manifold (the yellow line) and $M^T \Delta C' = [1.00, 1.62]$ meaning there are 1 ppb nitrogen atoms and 1.62 ppb oxygen atoms artificially added to the system. In this example, $M_{fix} \Delta C' \approx [1.97, 2.50, -2.50, 1.01, -2.21]$, correcting the mass imbalance to the green square in Figure 1.

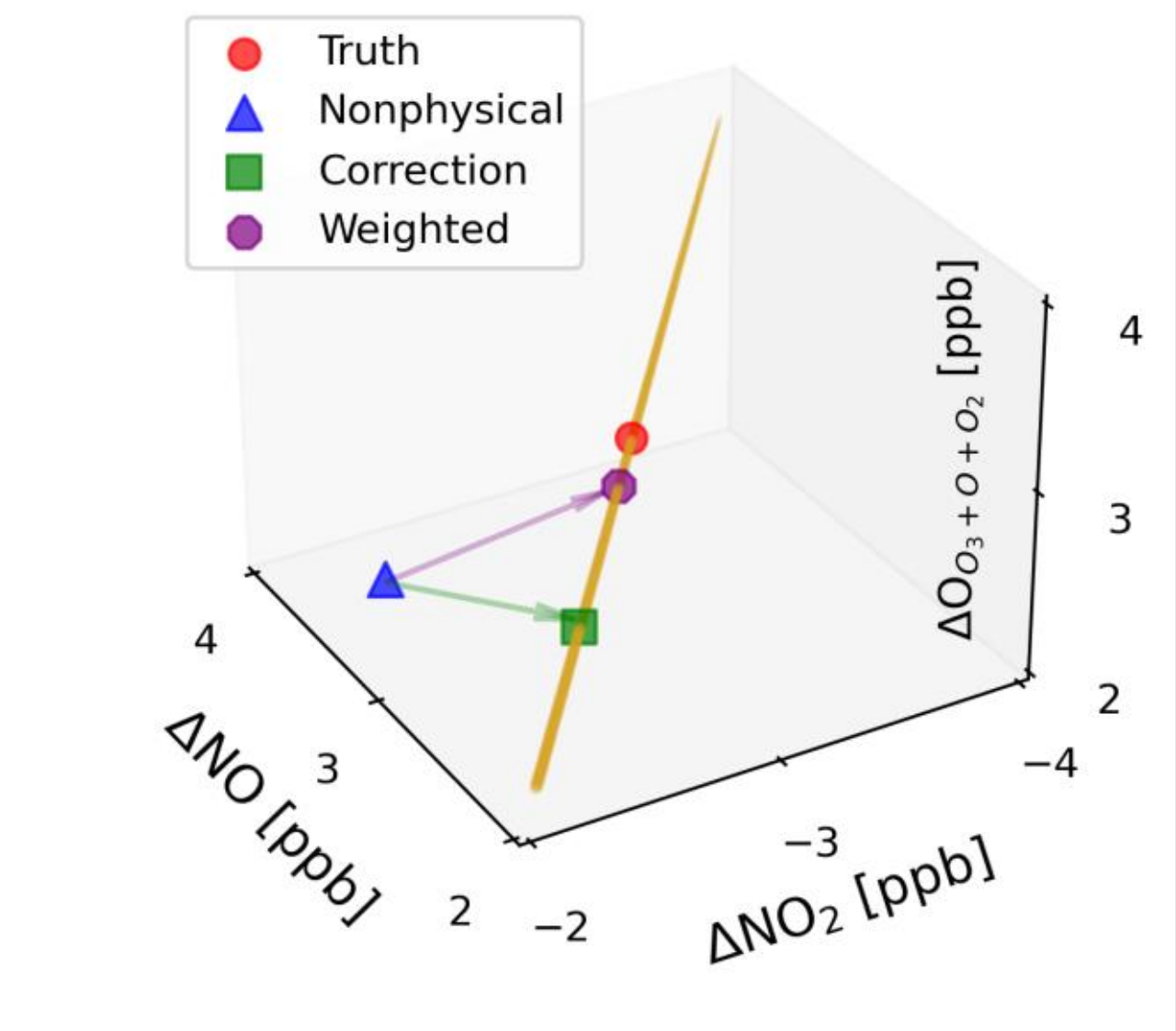

**Figure 1.** Visualization of the corrective approach for the primary photolytic cycle. The 3 physically consistent predictions (truth, correction, and weighted nudge) lie on a yellow line that represents the conservative manifold on which both nitrogen and oxygen atoms are conserved.

Let us now factor in uncertainty and assume that there is a spread of accuracy across variables on a test dataset, where we find NO is much better predicted than $NO_2$, and $O_3$ has better predictions than atomic or diatomic oxygen. We apply the uncertainty-weighted correction from Equation 5 to nudge the prediction to a more likely mass-conserving solution. The accuracies of this fictional test dataset yield $R^2$ values of 0.99 for $O_3$, 0.97 for NO, 0.92 for $NO_2$, 0.92 for O, and 0.92 for $O_2$, where we assume equal scaling factors in this example. In this case, $M_{fix,weighted}\Delta C' = [2.00, 2.84, -2.84, 0.78, -1.99]$ correcting the blue triangle to the purple octagon, which is now mass balancing (lying on the physically possible manifold). While not the closest mass-conserving solution, this weighted correction is much closer to the original solution because it nudged the species that were most uncertain, while leaving likely accurate species like NO and $O_3$ almost untouched.

## 2.4 The Julia photochemical mechanism

To evaluate this method in a real use case, we use the Julia photochemical model[11] as a reference model to train a machine learning XGBoost emulator. The Julia photochemical model is a simplified mechanism containing the reactions and species necessary for troposphere ozone photochemistry including $NO_x$ cycling, radical chemistry, and VOC oxidation. It has been used as a reference model for several data-driven and machine learning algorithms, including neural networks[19], sparse identification of nonlinear dynamics[36,37] and data-driven discovery of conservation laws[38].

Prior hard constraints work focused on carbon and nitrogen conservation using an older version of the Julia photochemical model which, without a carbon and nitrogen mixed-element species and ignoring hydrogen and oxygen balances, could be solved with two independent scaling factors. To demonstrate how this corrective approach automatically handles a mixed-element carbon and nitrogen species, for the rest of this work we use a version of the model augmented to include peroxyacetyl nitrate (PAN), which has the chemical formula $CH_3COO_2NO_2$. We update the Julia photochemical mechanism to include PAN and associated reactions, visualized in the mechanism graph in Figure 2. We include two important precursors of PAN, acetaldehyde and methylglyoxal[39], as well as the intermediate peroxyacetyl radical $CH_3C(O)OO$. We denote acetaldehyde, methylglyoxal, and peroxyacetyl radical as ALD2, MGLY, and $MCO_3$ respectively in our mechanism for consistency with the GEOS-Chem atmospheric chemistry module.

We include updates from Yang et al.[37] who further developed the Julia photochemical model to conserve all atoms, not just carbon and nitrogen, and construct it using the Julia package Catalyst.jl[40]. With this package, the time evolution of the chemical system can be simulated using the Julia library DifferentialEquations.jl[41], which includes a suite of stiffly-stable solvers. Besides the PAN-related expansion, another change between the new version of the model in the present work and the updates in Yang et al.[37] is that atomic oxygen and its reaction with diatomic oxygen is removed. This is a minor structural difference to the original mechanism in Sturm and Wexler[11], which by applying a pseudo-steady state assumption to atomic oxygen to reduce numerical stiffness, ensured that the atomic oxygen reaction proceeded at the same rate as $NO_2$ photolysis, effectively lumping the first two reactions together. We now represent these two reactions as a single reaction. We also ensure that none of the reaction rates include diatomic oxygen concentration as a factor in their rate laws, instead using the effective rate constant for termolecular reactions from the JPL 19-5 data evaluation "Chemical Kinetics and Photochemical Data for Use in Atmospheric Studies"[42]. We update all bimolecular and termolecular reactions to use rate constants recommended by JPL 19-5[42]. All photolytic rates are obtained using the KPP Standalone Interface within a GEOS-CF run[43], at a surface grid cell containing Los Angeles used in prior work exploring chemical cycling in the atmosphere[44].

We use the Julia photochemical model in the above configuration to simulate 1 million 1-hour runs, with tendencies reported every 5 minutes. As in prior work using this reference model, concentrations are randomly initialized within realistic ranges for every case[19,45]. We train the XGBoost algorithm to predict these 5-minute tendencies as targets, given species concentrations as input. In this multitarget regression problem, the tendency of each species is given its own XGBoost predictor, each of which is composed of 1000 estimators subsequently boosting the gradient with respect to mean squared error and a learning rate of 0.01 scaling the predictions of each estimator. Each estimator itself is a decision tree structure with a maximum depth of 10 levels. All other hyperparameters besides a random seed keyword argument are kept at package defaults (documentation available at https://xgboost.readthedocs.io/en/stable/, last access August 27, 2024). We use 90% of the cases for training and reserve 10% (100,000 1-hour runs) for testing in the following results section.

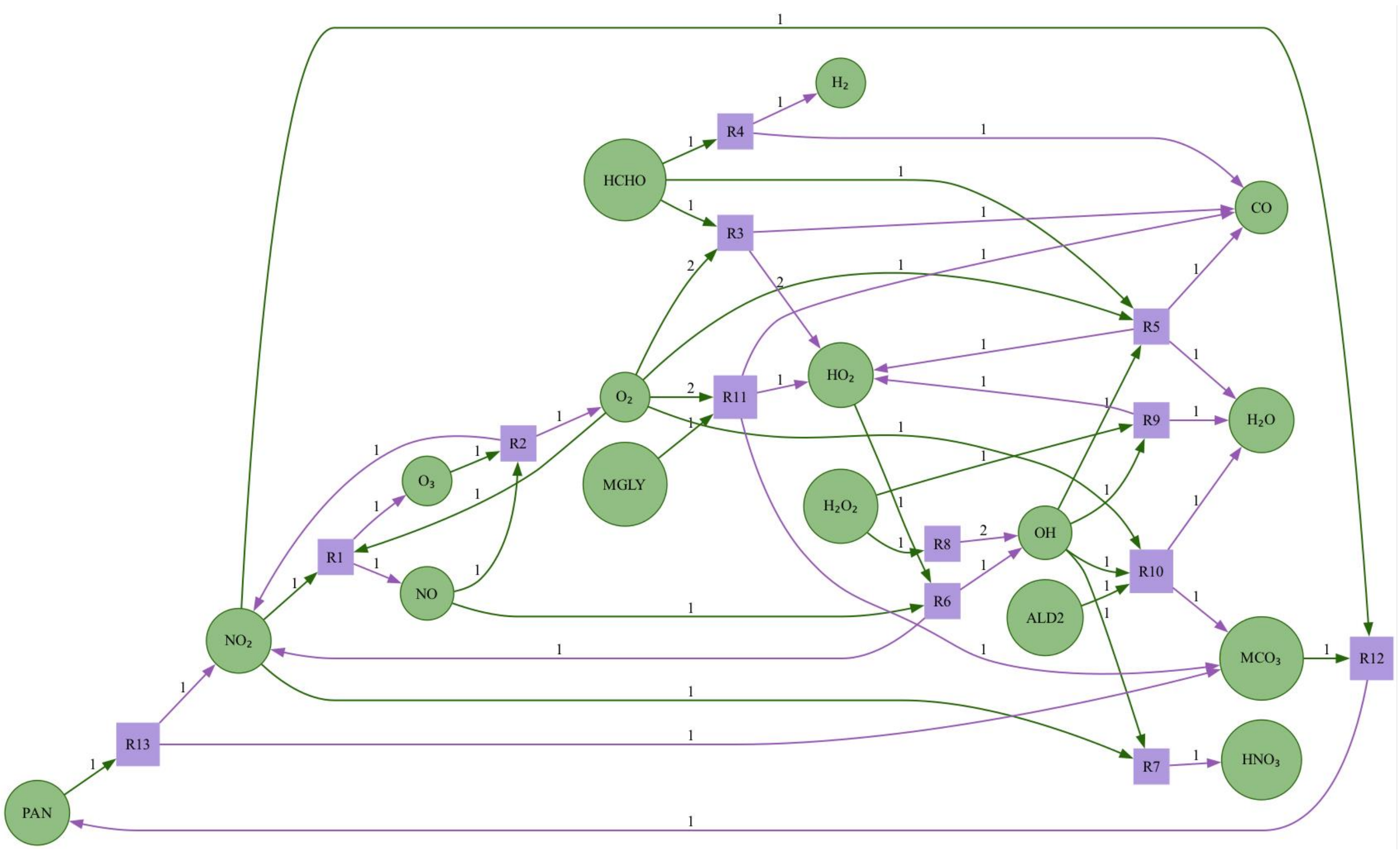

**Figure 2.** The updated Julia photochemical mechanism includes peroxyactyl nitrate (PAN) chemistry, as well as important precursors and associated radicals. The full reaction list is contained in S3 in the supplement.

# 3 RESULTS

## 3.1 Mass imbalance of XGBoost predictions

Before applying the corrective nudging approach, we first assess the extent to which atoms are not conserved in the XGBoost predictions. Figure 1 shows the distributions of the mass imbalances, based on the net tendencies (change in concentration over 5 minutes) of each element. The tendency using the corrective method is plotted as a vertical line at 0, as after nudging net atom tendencies are all 0 to machine precision.

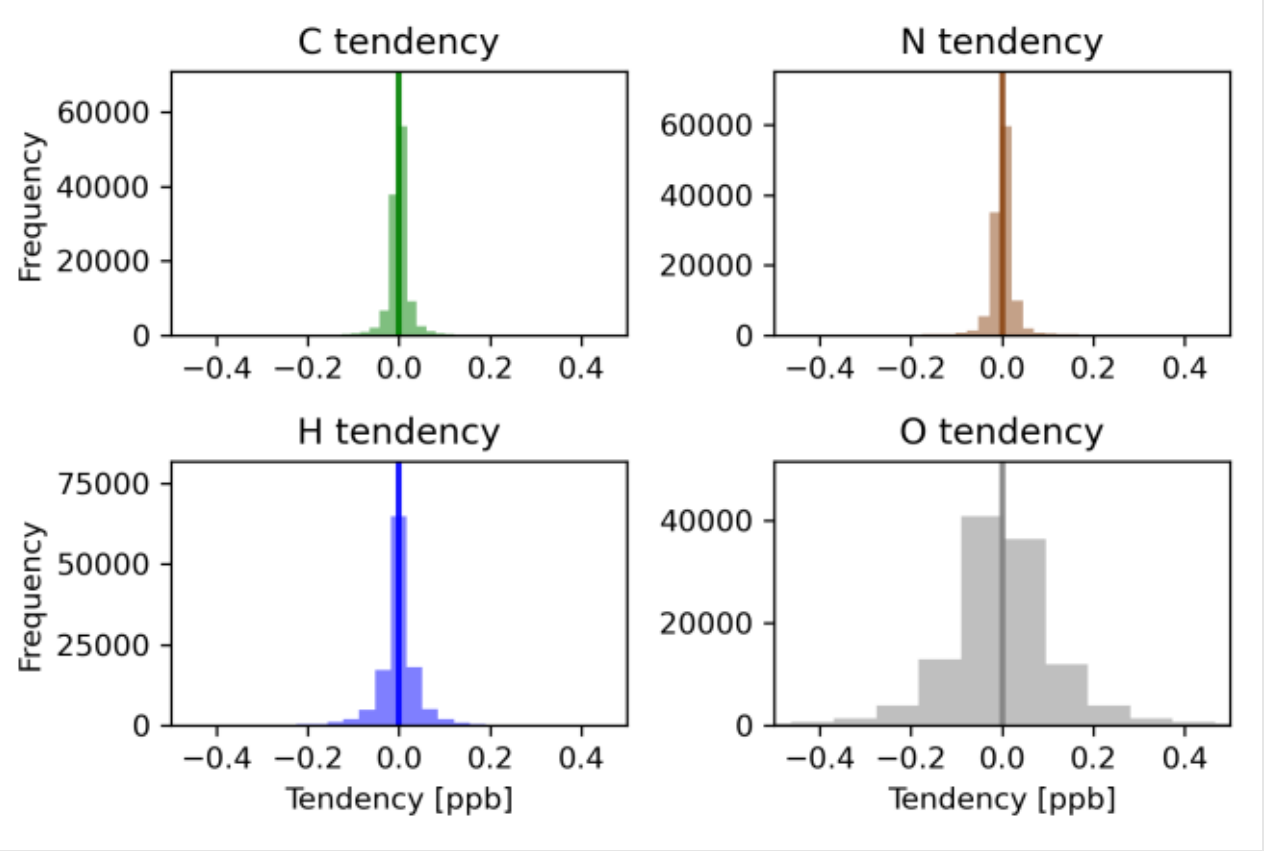

**Figure 3.** Distributions of the atom imbalance for each element.

For all elements, the distributions are near zero-centered, which indicates low bias (atoms are not, on average, added to or removed from the system in large amounts) and is consistent with the good performance of XGBoost for prediction in the atmospheric chemical sciences[23,46–48].

Though the single largest imbalance is a contribution of 4.92 ppb over 5 minutes, the majority of all the distributions remains within smaller bounds on the order of hundreds of parts per trillion, which is ca. 10% the size of 5-minute tendencies for many of the species. This underscores that these nonphysical deviations from machine learning models can often be minor, in which case the predictions only need a small nudge to a solution that conserves mass.

## 3.2 Effect of nudging on accuracy

In line with the relatively minor nonphysical atom conservation deviations shown in section 3.1, the XGBoost ensemble is able to predict the tendencies of all individual species well, with high $R^2$ and low error for all species (shown in S4 in the supplement). The least accurately predicted species is OH, where performance statistics are still reasonably good ($R^2$=0.9707, RMSE=$9\times10^{-5}$ ppb). This result allows us to ask a targeted question: does the corrective nudging approach decrease the quality of predictions, or phrased differently, does this approach ensure mass conservation at the cost of accuracy? The following results indicate that the unweighted nudge does indeed sacrifice accuracy for physical consistency, while the weighted nudge actually improves accuracy slightly while transforming the predictions to conserve mass.

Figure 4 shows the accuracy of predictions for four representative species: $O_3$, NO, OH, and PAN. The uncorrected prediction accuracy with respect to the reference model is shown on the first row, the nudging approach on the second row, and the species-level weighted approach on the third row. The carbon and nitrogen mixed species PAN is already very well predicted before applying constraints and only receives a slight accuracy loss with the unweighted correction in row two, largely stemming from perturbations to smaller tendencies. The very accurate predictions of $O_3$ and NO are barely changed by the adjustment in the second row, though slightly improved.

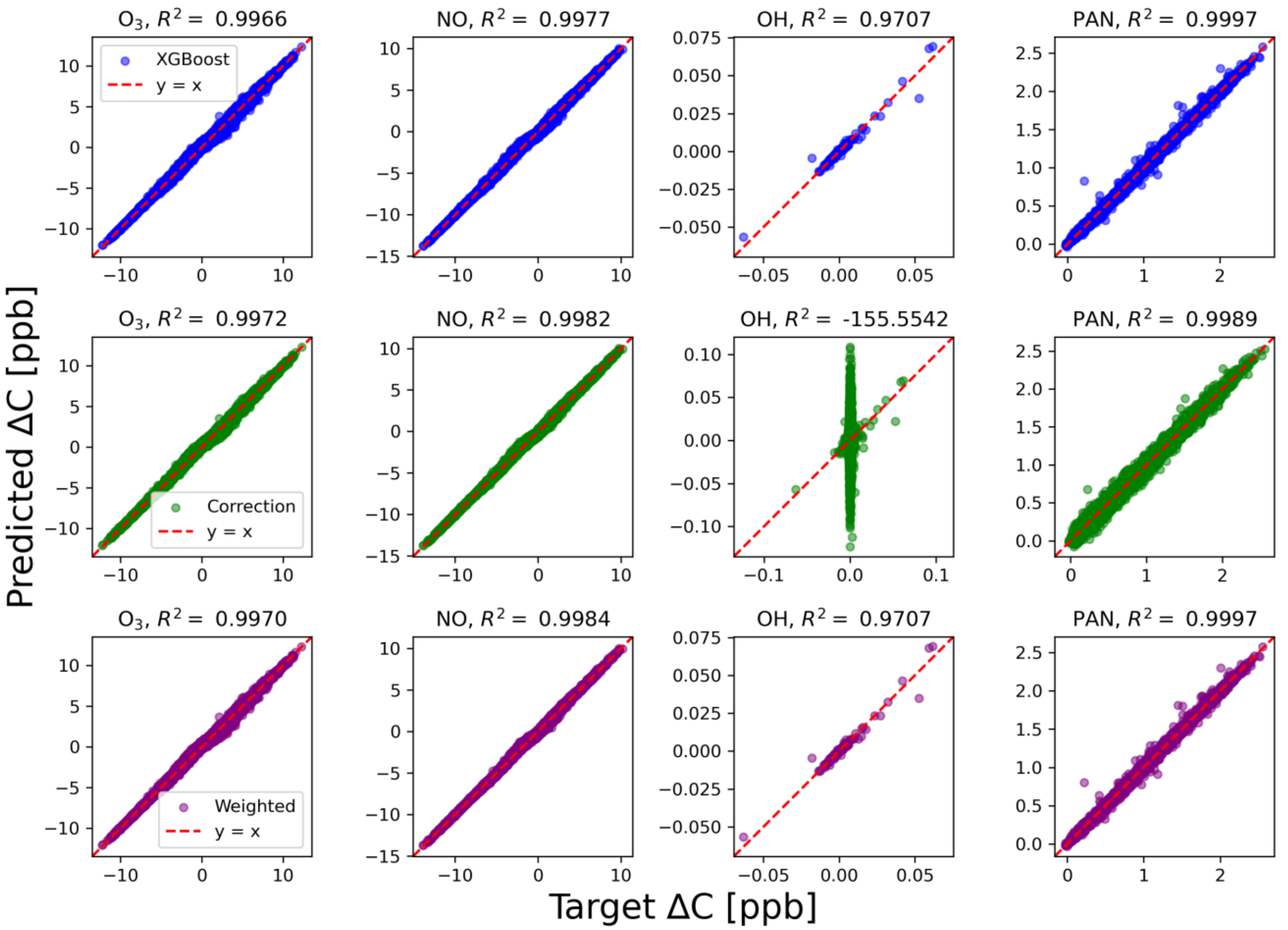

**Figure 4.** Scatter plot of key species with the uncorrected predictions, corrected predictions and the weighted correction.

The nudging correction in the second row obliterates the accuracy of OH predictions, which have much smaller absolute tendencies. The difficulty of capturing OH was also seen in prior work using hard-constrained deep learning embedded with the mass-conserving framework[19]. As in previous work, we could design the approach for this specific system to neglect conservation of oxygen and hydrogen, which have diatomic oxygen and water as extremely large source/sink terms; in that case, we could just remove columns of $M$ and leave OH unperturbed. However, we seek a more general framework that can handle radicals, as some applications may have radicals or other important species that are mixtures of multiple elements that need to be conserved: for example, reservoir chlorine nitrate is an important sink and source of reactive chlorine and nitrate which are in turn sinks of stratospheric ozone[49–51]. The performance of OH in the second row indicates that the unweighted nudging approach may not be appropriate for predictions that span orders of magnitude across different variables. We turn to the species-level weighting approach to address this scaling issue.

We design the species-level weighting approach and corresponding weighting function to be able to handle predictions of small radicals. The third row of Figure 4 shows that OH accuracy is not adversely affected when using the species-weighted nudging approach. That is because the species-level weighting approach contains two factors, a scaling factor and uncertainty factor. The first factor is simply a scaling factor to account for the magnitude-spanning spread of tendencies across different species. This scaling factor in the species-level weighting allows for small radicals to be well predicted after the nudge, in this case OH.

The second factor represents the uncertainty in species predictions as inversely related to $R^2$ of the uncorrected approach on a test data set. This uncertainty factor, with weights appropriately chosen, encourages the global solution to be more accurate. The overall $R^2$ of all species besides OH is 0.9984 for the uncorrected predictions and the unweighted approach degrades predictions during the nudge to an $R^2$ of 0.9905. In contrast, the weighted approach yield improves predictions slightly during the nudge to an $R^2$ of 0.9986. This is expected behavior and happens by design: rather than nudging to whatever is the closest prediction (from a least squares perspective) that conserves mass, this approach considers the uncertainties of different species to nudge to a more likely mass-conserving solution.

### 3.3 Further exploration of species-level weighting

The species-level weights improve both OH radical prediction and the global metrics of the solution. Figure 5 visualizes these weights, which are comprised of both a scaling factor and an uncertainty factor. We note that the 3 most heavily weighted species are the radicals OH, $MCO_3$, and $HO_2$. Though other approaches to species-level weighting are possible, including manual selection of importance, these results show Equation 6 is appropriate for the photochemical system investigated here, automatically weighing both scale and uncertainty without requiring any manual selection.

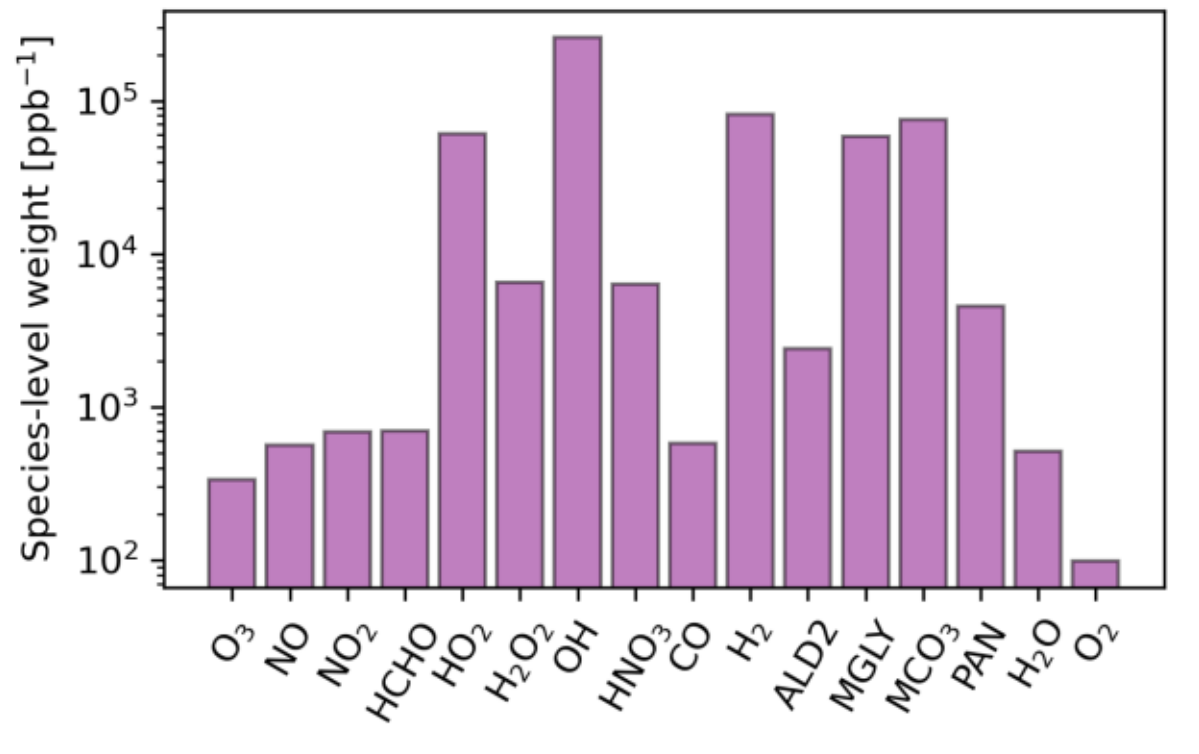

**Figure 5.** Species-level weights used in the weighted correction.

Figure 6 visualizes the parts of the projection matrices that determine the perturbation to $\Delta C'$. This is the non-identity part of the correction matrix in Equations 3 and 5 for the unweighted and the weighted approaches respectively (for visual purposes, the identity component adding 1 to the diagonals is omitted). The unweighted matrix is not data-driven, but rather a static property emerging from the molecular formulae in the system. The weighted matrix is data-driven in that it depends on the test data uncertainty in predictions and average absolute tendencies.

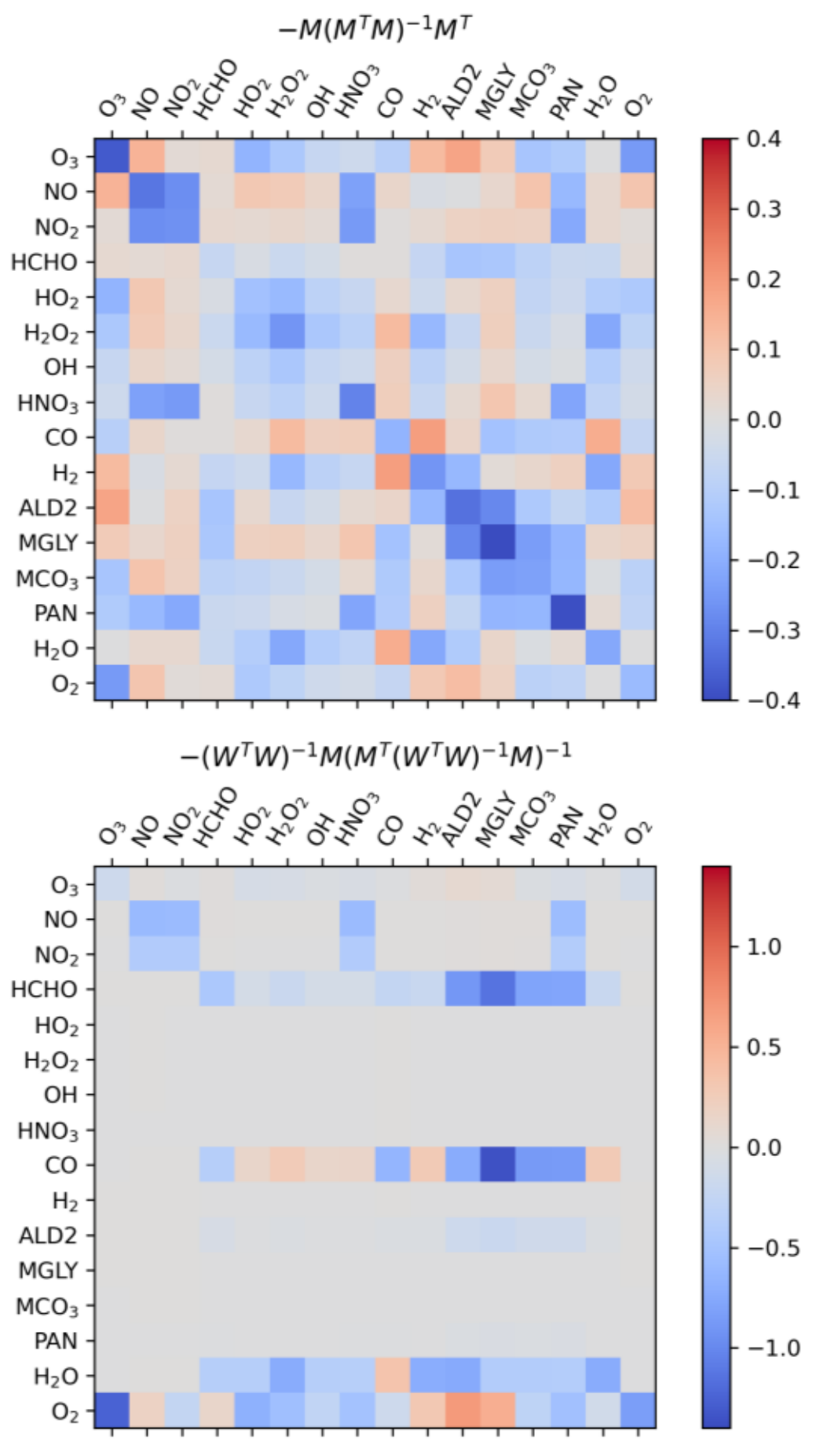

**Figure 6.** The term of the projection matrices determining to the perturbation to $\Delta C'$. Note that these are added to the identity terms to create the nudging projection matrices $M_{fix}$ and $M_{fix,weighted}$.

The unweighted projection is fully symmetric, as is the perturbation component of it (within a relative tolerance of $10^{-13}$). The weighted approach is asymmetric and sparse. Species with high weights have essentially no perturbation, for example the OH row for the weighted matrix. This reflects that the choice of species-level weighting is automatically more targeted in the species that it nudges.

## CONCLUSIONS

If not anchored by hard constraints, numerical predictions of atmospheric chemical behavior will yield nonphysical results. This includes not conserving mass, or not conserving atoms in a machine learned chemical reaction network; models based on data rather than fundamental process knowledge are particularly prone to such nonphysical results. These nonphysical violations can be minor, requiring only a small adjustment. Our corrective approach based on least-squares finds the smallest possible nudge to the predictions that guarantees conservation of atoms to machine precision.

When tested on an XGBoost-predicted ozone photochemical system, where the changes of different species span orders of magnitude, we find that species-level weights are essential for an accurate correction. A scaling factor needs to be introduced to accurately model hydroxyl radical, an important species with low absolute concentrations. An uncertainty factor in the species-level weights improves the overall accuracy in predictions. Rather than making the smallest possible adjustment to the predictions, the weighted approach uses species-level uncertainty to target which species to adjust. For the photochemical system studied here, this leads to a moderate improvement in overall accuracy.

The atom-conserving nudge can act as a wrapper for any numerical approach and does not require custom architectures[19] or training data[11] as in prior efforts. Included in this wrapper can be a calculation of tendencies before subsequent correction if the predictions are concentrations. Only the elements of interest need to be constrained such that their atoms are conserved, shown in the comparison of corrective vs. flux-based approaches in S5 in the supporting information. For this reason, the approach is well-suited as an automated unit test to test mass closure in larger chemical mechanisms and could also be paired with an automatic fix if the numerical solvers do violate conservation of atoms. Beyond numerical models, this wrapper has potential application to experimental chamber chemistry data analogously to the approach in Brown[34] that improved model-measurement agreement when correcting applying a least-squares mass correction in a raindrop collision application. For example, empirical data often have closure issues from processes like wall losses[52] and protocols have been developed to standardize chamber wall loss characterization experiments[53].

This uncertainty-weighted nudging approach is not limited to being used as a corrective wrapper: it could be built into a neural network layer as another option for automatic hard constraints in deep learning, alongside prior approaches in the Earth system sciences[6–8,19] and more generally in the physical sciences[12,16,17,21,54–56], as well as an alternative way of representing and handling uncertainty in deep learning[57–59]. Potential architectures this atom-conserving nudging approach could be built into are deep learning autoencoders integrating a chemical mechanism forward as in Kelp et al.[45,60] but bounded in a stoichiometrically resolved and mass-conserving latent space, or graph neural networks mapping between low and high chemical complexity while conserving the total budget of atoms.

Beyond atom conservation, the uncertainty-weighted approach could be used in other applications combining variable-level uncertainty with hard constraints to nudge predictions to be both physically consistent and more accurate.

## ASSOCIATED CONTENT

### Data Availability Statement

The exact versions of the scripts and model output used for analysis and figures in this work are available at https://doi.org/10.5281/zenodo.13385987; the exact version of the Julia photochemical model used to produce the model output, as well as the modifications to Catalyst.jl for Figure 3, is available at https://doi.org/10.5281/zenodo.13385541.

### Supporting Information

The supporting information (PDF) contains the following sections:
S1 Derivation of the weighted approach
S2 Modification for concentrations instead of tendencies
S3 Julia photochemical model reaction list
S4 Scatter plots of accuracy for all other species
S5 Comparison to another hard constraints framework

## AUTHOR INFORMATION

### Corresponding Author

Obin Sturm, psturm@usc.edu

### Author Contributions

Both authors wrote and have given approval to the current version of the manuscript.

## ACKNOWLEDGMENT

P.O. Sturm acknowledges funding from NASA Grant Number 80NSSC23K0523. The authors would like to thank Paula Harder, Tony Wexler, and Chris Tessum for valuable discussion.

## ABBREVIATIONS

ALD2, acetaldehyde; GEOS, Goddard Earth Observing System; GEOS-CF, Goddard Earth Observing System Composition Forecasts; MCO3, peroxyacetyl radical; MGLY, methylglyoxal; PAN, peroxy acetyl nitrate; ReLU, rectified linear unit; XGBoost, extreme gradient boosting

Supporting Information for

# *A nudge to the truth: atom conservation as a hard constraint in models of atmospheric composition using an uncertainty-weighted correction*

Patrick Obin Sturm (psturm@usc.edu) and Sam J. Silva

## S1. Derivation of the weighted correction approach using Lagrange multipliers

We want to find the best nudge to a set of concentration changes of chemical species such that atoms in the system are conserved exactly.

Mathematically, we seek an optimal correction $\Delta C \in \mathbb{R}^m$ to a set of $m$ species concentrations $\Delta C' \in \mathbb{R}^m$, where the species are formed by atomic combinations of $p < m$ elements in $M \in \mathbb{Z}_{\geq 0}^{m,p}$ such that no atoms from any of the elements are added or removed in a nonphysical way. To keep it general, we assign species-level weights $W \in \mathbb{R}_{\geq 0}^{m,m}$ which determine the importance of each species; important species are adjusted less. In the case that $W$ is the same for all species, then this is an unweighted approach that finds the smallest correction.

This can be framed as a constrained optimization problem:

$$min_{\Delta C} \|W(\Delta C' - \Delta C)\|_2 \quad \text{S1.1}$$

Subject to the constraints

$$M^T \Delta C = 0_p \quad \text{S1.2}$$

We note that other constraints could be added, e.g. element-wise inequality constraints, $-M_{fix}\Delta C' < C$ for all $\Delta C' < 0_m$, but neglect these in order to be able to obtain a closed-form solution. We also note that other measures of distance besides the L2-norm exist but use this norm for the same purpose of obtaining a closed-form solution. We use the method of Lagrange multipliers to obtain the solution to this constrained least squares problem[1], beginning with setting up the Lagrangian function:

$$\mathcal{L}(\Delta C, \lambda) = (\Delta C' - \Delta C)^T W^T W(\Delta C' - \Delta C) + \lambda^T (M^T \Delta C) \quad \text{S1.3}$$

Where the Lagrangian $\mathcal{L}$ is a scalar function that combines the original objective function (the first term is S1.1 squared) with the constraints as a second term, weighted by the Lagrangian multiplier $\lambda \in \mathbb{R}^p$.

We seek the $\Delta C$ where the derivative of the Lagrangian function with respect to $\Delta C$ is zero:

$$\frac{\partial \mathcal{L}}{\partial \Delta C} = 2W^T W(\Delta C' - \Delta C) + M\lambda = 0_m \quad \text{S1.4}$$

that simultaneously obeys the constraint in S1.2, which holds where the derivative of the Lagrangian function with respect to $\lambda$ is zero:

$$\frac{\partial \mathcal{L}}{\partial \lambda} = M^T \Delta C = 0_p \quad \text{S1.5}$$

We now use these two equations to solve for the optimal, atom-conserving $\Delta C$. We rearrange S1.4 to isolate $\Delta C - \Delta C'$:

$$\frac{1}{2}(W^TW)^{-1}M\lambda = \Delta C' - \Delta C \qquad \text{S1.6}$$

Then, we multiply both sides by $M^T$, keeping in mind that $M^T\Delta C = 0_p$:

$$\frac{1}{2}M^T(W^TW)^{-1}M\lambda = M^T\Delta C' - \cancel{M^T\Delta C} \qquad \text{S1.7}$$

Allowing us to isolate $\lambda$:

$$\lambda = 2(M^T(W^TW)^{-1}M)^{-1}M^T\Delta C' \qquad \text{S1.8}$$

We can substitute $\lambda$ for equation S1.8 in equation S1.4:

$$\frac{\partial \mathcal{L}}{\partial \Delta C} = 2W^TW(\Delta C' - \Delta C) + 2M(M^T(W^TW)^{-1}M)^{-1}M^T\Delta C' = 0_m \qquad \text{S1.9}$$

Dividing by 2 and isolating the term with $\Delta C$, we get

$$W^TW\Delta C = W^TW\Delta C' - M(M^T(W^TW)^{-1}M)^{-1}M^T\Delta C' \qquad \text{S1.10}$$

To isolate the optimal nudge $\Delta C$, we multiply both sides by $(W^TW)^{-1}$:

$$\Delta C = \Delta C' - (W^TW)^{-1}M(M^T(W^TW)^{-1}M)^{-1}M^T\Delta C' \qquad \text{S1.11}$$

And to get this in the form of a single matrix multiplication, we factor $\Delta C'$ out of the right-hand side to obtain

$$\Delta C = (I - (W^TW)^{-1}M(M^T(W^TW)^{-1}M)^{-1}M^T)\Delta C' \qquad \text{S1.12}$$

Where $I$ is the identity matrix of size $m$ and we define the weighted correction matrix as in section 2.2 in the main text:

$$M_{fix,weighted} = I - (W^TW)^{-1}M(M^T(W^TW)^{-1}M)^{-1} \qquad \text{S1.13}$$

We note that if we multiply S1.11 by $M^T$, the right-hand side becomes $M^T\Delta C' - M^T\Delta C' = 0_p$, satisfying S1.2 to conserve atoms. We also note that if the weight matrix is equal to identity or a diagonal matrix with equal elements, or if it is orthonormal, the problem simplifies to the unweighted correction in section 2.1 of the main text.

## S2. Nudging concentrations to conserve total amounts

This approach could be augmented to nudge total concentrations, if concentrations instead of tendencies are predicted. To find the smallest perturbed concentration mathematically, we seek

$$min_C \|C' - C\|_2 \qquad \text{S2.1}$$

subject to

$$M^T C = A \qquad \text{S2.2}$$

where $C' \in \mathbb{R}_{\geq 0}^{m,1}$ is a prediction of $m$ species chemical concentrations, $C \in \mathbb{R}_{\geq 0}^{m,1}$ is the corrected prediction, and $A \in \mathbb{R}_{\geq 0}^{p,1}$ is a vector containing the total atom concentrations of $p < m$ elements, which is the constraint that needs to be satisfied. $M \in \mathbb{Z}_{\geq 0}^{m,p}$ is the same composition matrix as in Section S1 and Section 2.1 in the main text.

The optimal $C$ (optimal with respect to S2.1) that satisfies S2.2 is

$$C = C' + M(M^T M)^{-1}(A - M^T C') \qquad \text{S2.3}$$

There are different strengths and weaknesses of adjusting concentrations versus tendencies. The tendencies approach has the advantage of $\Delta C$ not necessarily needing to remain positive. In addition, incremental changes are less likely to span many orders of magnitude. However, the tendencies or prior concentrations to calculate may not always be known in every application: an example of this is mapping back and forth between latent dimensionality and full dimensionality[2–5] or conserving a set of properties when converting between lumped superspecies and species[5].

### S3. Julia photochemical mechanism

| Table S1. Species | |
|---|---|
| Name | Species ID |
| Ozone | $O_3$ |
| Nitric oxide | NO |
| Nitrogen dioxide | $NO_2$ |
| Formaldehyde | HCHO |
| Hydroperoxyl radical | $HO_2\cdot$ |
| Hydrogen peroxide | $HO_2H$ |
| Hydroxyl radical | $OH\cdot$ |
| Nitric acid | $HNO_3$ |
| Carbon monoxide | CO |
| Diatomic hydrogen | $H_2$ |
| Acetaldehyde | ALD2 |
| Methylglyoxal | MGLY |
| Peroxyacetyl radical | $MCO_3$ |
| Peroxyacetyl nitrate | PAN |
| Water | $H_2O$ |
| Diatomic oxygen | $O_2$ |

The set of species in Table S1 corresponds to the following composition matrix $M$:

$$M = \begin{bmatrix} species & \mathrm{C} & \mathrm{N} & \mathrm{H} & \mathrm{O} \\ \mathrm{O_3} & 0 & 0 & 0 & 3 \\ \mathrm{NO} & 0 & 1 & 0 & 1 \\ \mathrm{NO_2} & 0 & 1 & 0 & 2 \\ \mathrm{HCHO} & 1 & 0 & 2 & 1 \\ \mathrm{HO_2} & 0 & 0 & 1 & 2 \\ \mathrm{H_2O_2} & 0 & 0 & 2 & 2 \\ \mathrm{OH} & 0 & 0 & 1 & 1 \\ \mathrm{HNO_3} & 0 & 1 & 1 & 3 \\ \mathrm{CO} & 1 & 0 & 0 & 1 \\ \mathrm{H_2} & 0 & 0 & 2 & 0 \\ \mathrm{ALD2} & 2 & 0 & 4 & 1 \\ \mathrm{MGLY} & 3 & 0 & 4 & 2 \\ \mathrm{MCO_3} & 2 & 0 & 3 & 3 \\ \mathrm{PAN} & 2 & 1 & 3 & 5 \\ \mathrm{H_2O} & 0 & 0 & 2 & 1 \\ \mathrm{O_2} & 0 & 0 & 0 & 2 \end{bmatrix}$$

| Table S2: Julia photochemical mechanism | | |
|---|---|---|
| Reaction Number | Reaction | Rate constant reference |
| R1 | $NO_2 + O_2 + hv \rightarrow NO + O_3$ | See note 1 |
| R2 | $O_3 + NO \rightarrow NO_2 + O_2$ | JPL 19-5[6] page 84 Table 1C |
| R3 | $HCHO + 2O_2 + hv \rightarrow 2\ HO_2^{\cdot} + CO$ | See note 1 |
| R4 | $HCHO + hv \rightarrow H2 + CO$ | See note 1 |
| R5 | $HCHO + HO^{\cdot} \rightarrow HO_2^{\cdot} + CO + H_2O$ | JPL 19-5[6] page 106 Table 1D |
| R6 | $HO_2^{\cdot} + NO \rightarrow OH^{\cdot} + NO_2$ | JPL 19-5[6] page 83 Table 1C |
| R7 | $OH^{\cdot} + NO_2 \rightarrow HNO_3$ | See note 2, JPL 19-5[6] page 434 Table 2-1 |
| R8 | $HO_2H + hv \rightarrow 2\ OH^{\cdot}$ | See note 1 |
| R9 | $HO_2H + OH^{\cdot} \rightarrow H_2O + HO_2^{\cdot}$ | T-independent recommendation, JPL 19-5[6] page 73 |
| R10 | $ALD2 + OH + O_2 \rightarrow MCO_3 + H_2O$ | JPL 19-5[6] page 107 Table 1D |
| R11 | $MGLY + 2O_2 \rightarrow MCO_3 + CO + HO_2^{\cdot}$ | See note 1 |
| R12 | $MCO_3 + NO_2 \rightarrow PAN$ | See note 2, JPL 19-5[6] page 435 Table 2-1 |
| R13 | $PAN \rightarrow MCO_3 + NO_2$ | Equil. Constant / k12, JPL 19-5[6] page 521 Table 3-1 |

Note 1: Photolysis rate taken from the GEOS-CF[7] full chemical state sampled using the KPP Standalone Interface at a grid cell over Los Angeles at local noon[8]

Note 2: Termolecular reaction with effective second-order rate constant calculated from low pressure and high pressure limits and total gas concentration[6]

## S4. Scatter plots for all other species

This section contains scatter plots of accuracy for the predictions as in Section 3.2 in the main text, for all other species in the Julia photochemical mechanism.

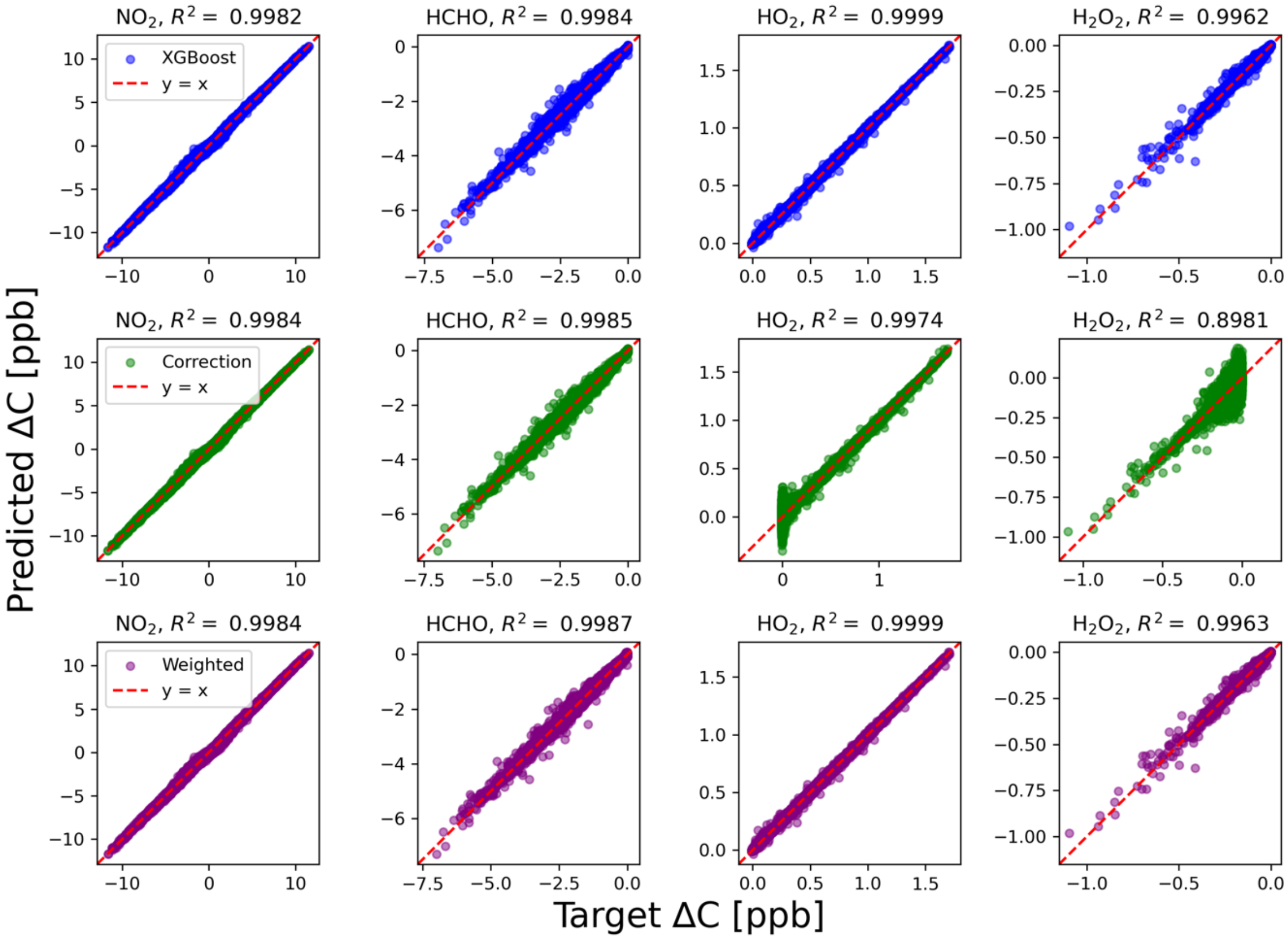

**Figure S1.** Scatter plot of $NO_2$, HCHO, $HO_2$, and $H_2O_2$ with the uncorrected predictions, corrected predictions and the weighted correction.

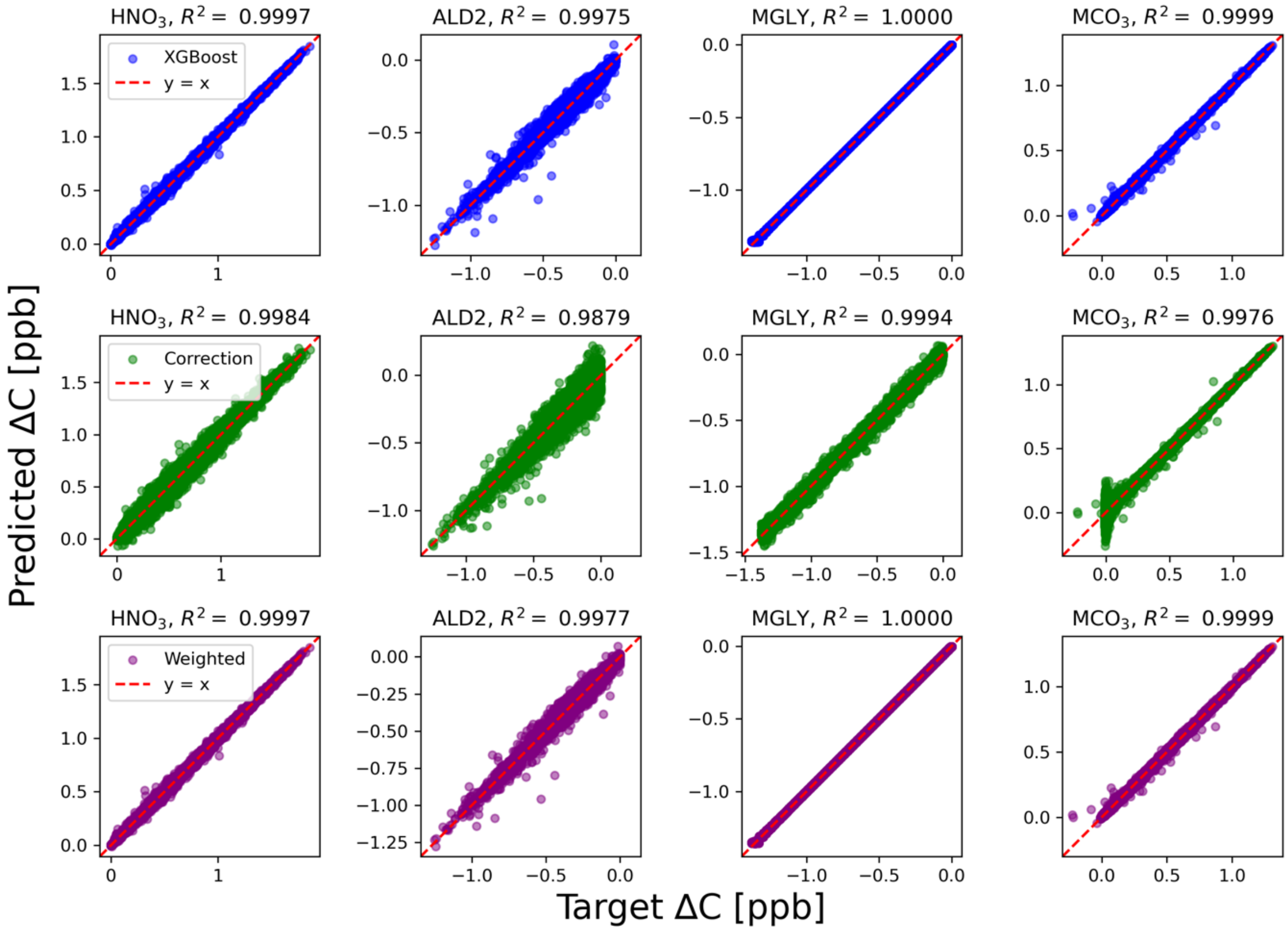

**Figure S2.** Scatter plot of $HNO_3$, ALD2, MGLY, and $MCO_3$ with the uncorrected predictions, corrected predictions and the weighted correction.

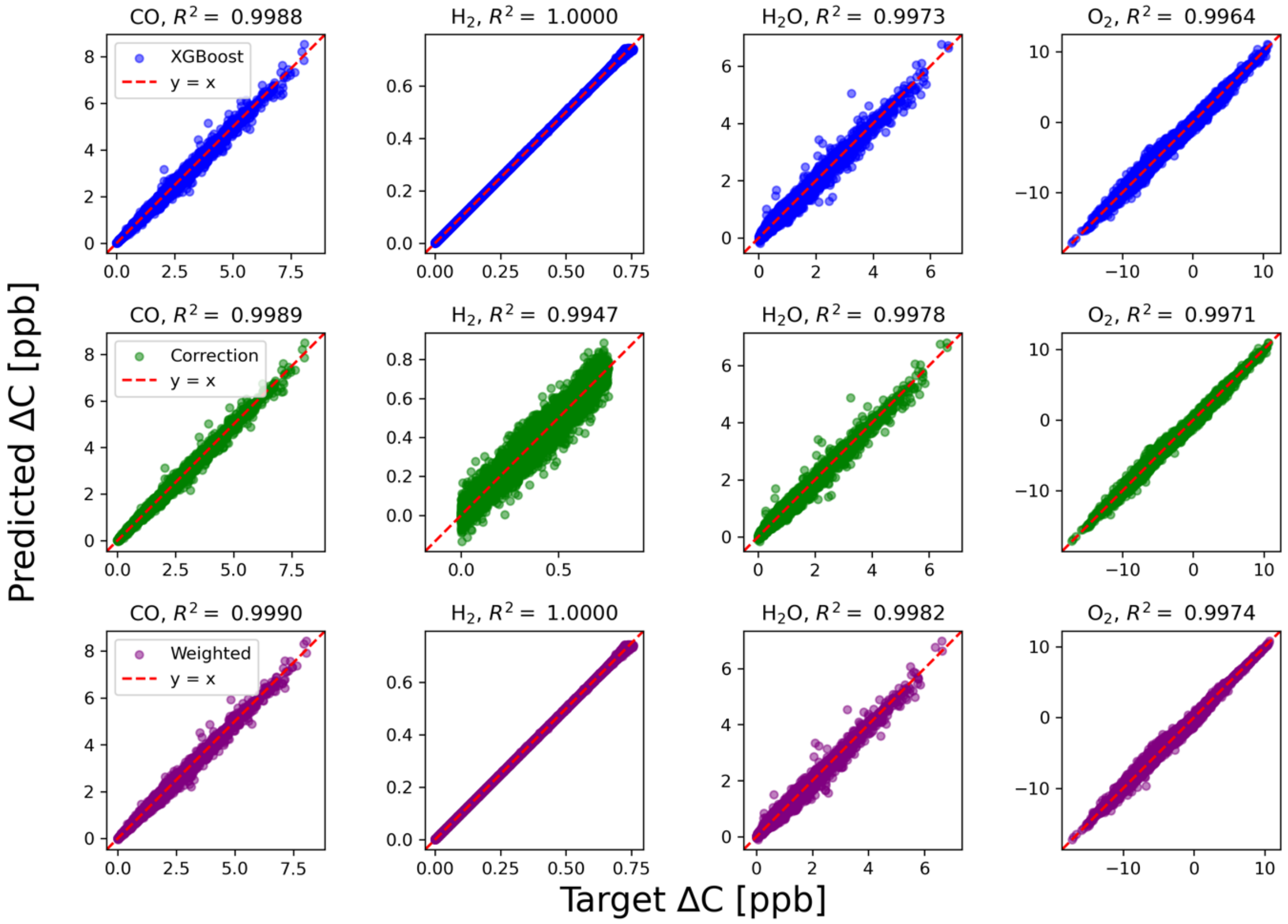

**Figure S3.** Scatter plot of CO, $H_2$, $H_2O$, and $O_2$ with the uncorrected predictions, corrected predictions and the weighted correction.

## S5. Comparison to the flux-based approach for previously trained neural networks

We can demonstrate these methods on prior neural networks (NN) trained to emulate a prior version of the Julia photochemical model[9,10]. The version of the Julia photochemical model used here contained 11 species and 10 reactions. In this work, conservation of carbon and nitrogen atoms was the focus: oxygen atoms were not conserved, as diatomic oxygen was treated as an infinite source and sink, and hydrogen atoms were not conserved, as water was not tracked by the model. We demonstrate here that the nudging approach can be used to just conserve a subset of atoms.

This prior work compared a physics-constrained NN embedded with the mass-conserving framework to a more standard architecture "intermediate" complexity NN without physical constraints but with a similar number of trainable hyperparameters. We apply this post-prediction correction to the output of this intermediate NN. Figure S4 shows that this has only a minor effect on accuracy of 4 key species (ozone, NOx, and formaldehyde) relative to the prediction of the Julia photochemical model. For $NO_2$, $R^2$ value of tendencies even agrees more closely with the Julia photochemical model after the atom-conserving correction is applied to the intermediate NN, reaching 0.95, which is the same as the physics-constrained NN embedded with the graph locality of the chemical mechanism. Numerical tests over all test data show that total carbon is conserved in all cases within $10^{-17}$ ppb and total nitrogen within $10^{-15}$ ppb.

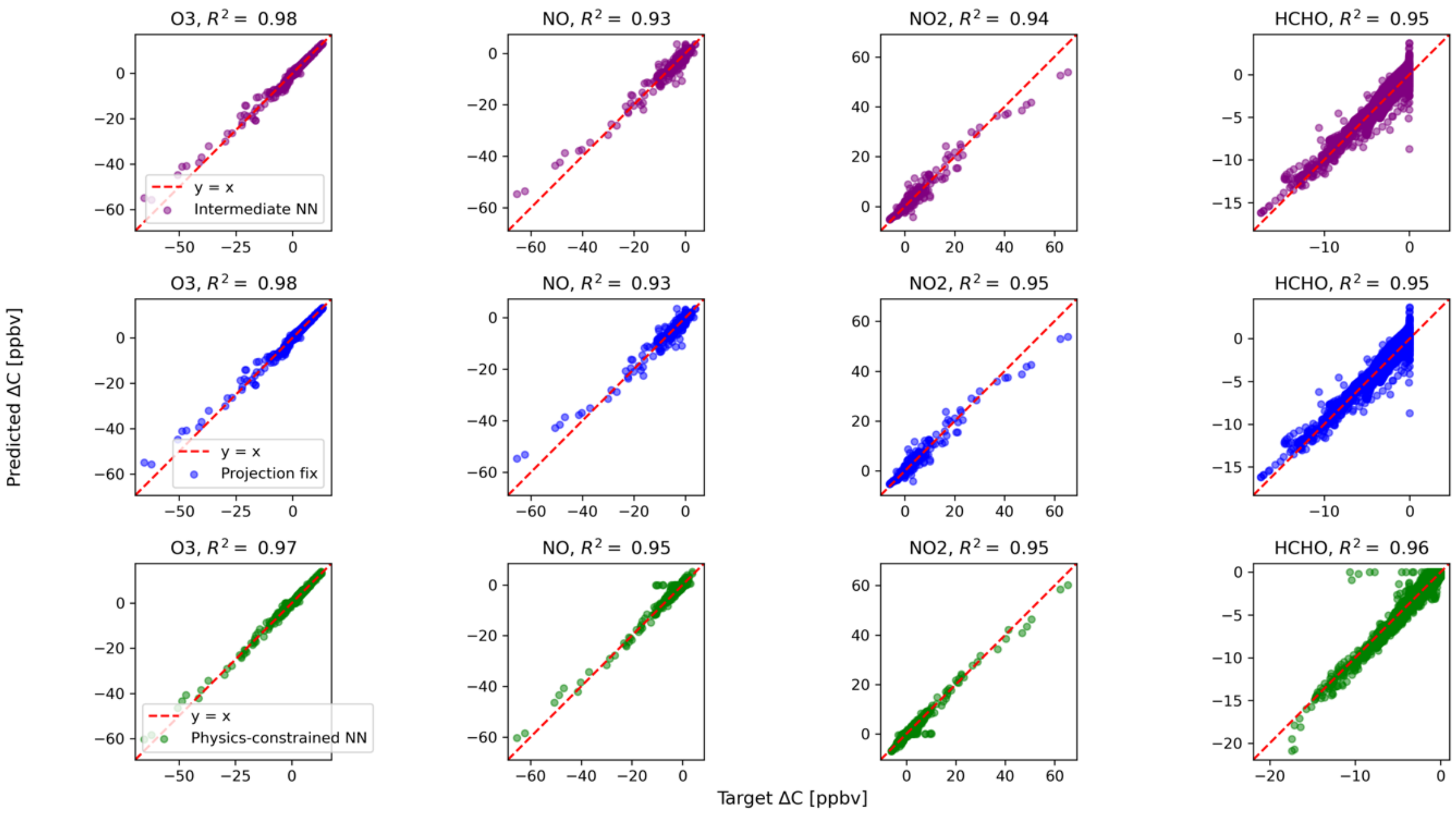

**Figure S4.** Scatter plot of target vs predicted tendencies, for the intermediate NN, the intermediate NN with the atom-conserving tendency correction using the methods from Section 2.1, and the physics-constrained NN from Sturm and Wexler[10].